\documentclass[prb,aps,showkeys,twocolumn,showpacs,amsmath,amssymb]{revtex4}

\DeclareMathOperator{\trace}{tr}
\newcommand* {\etal}{\emph{et~al.}}
\newcommand* {\vek}{\bm}
\newcommand* {\dd}{\mathrm{d}}
\newcommand* {\ee}{\mathrm{e}}
\newcommand* {\fermi}{\mathrm{F}}
\newcommand* {\kbolz}{k_\mathrm{B}}

\usepackage{graphicx}
\usepackage{bm}
\usepackage{array}

\begin{document}

\title{Temperature dependence of D'yakonov-Perel' spin relaxation
in zinc blende semiconductor quantum structures}

\author{J.~Kainz}
\email{josef.kainz@physik.uni-regensburg.de}
\author{U.~R\"ossler}
\affiliation{
Institut f\"ur Theoretische Physik, Universit\"at Regensburg,
93040 Regensburg, Germany}

\author{R.~Winkler}
\affiliation{
Institut f\"ur Festk\"orperphysik,
Universit\"at Hannover,
Appelstr.~2,
30167 Hannover,
Germany}

\date{July 14, 2004}

\begin{abstract}
  The D'yakonov-Perel' mechanism, intimately related to the spin
  splitting of the electronic states, usually dominates the spin
  relaxation in zinc blende semiconductor quantum structures.
  Previously it has been formulated for the two limiting cases of
  low and high temperatures. Here we extend the theory to give an
  accurate description of the intermediate regime which is often
  relevant for room temperature experiments.  Employing the
  self-consistent multiband envelope function approach, we determine
  the spin splitting of electron subbands in $n$-(001) zinc blende
  semiconductor quantum structures.  Using these results we
  calculate spin relaxation rates as a function of temperature and
  obtain excellent agreement with experimental data.
\end{abstract}

\maketitle

%%%%%%%%%%%%%%%%%%%%%%%%%%%%%%%%%%%%%%%%%%%%%%%%%%%%%%%%%%%%%%%%%%%%
\section{Introduction}

In the context of spintronic devices, the spin degree of freedom has
recently attracted considerable interest. \cite{divincenzo:1995,
prinz:1998, wol01, aws02, oes02} Long spin-relaxation times are
crucial for the operation of such devices. Therefore, a quantitative
understanding is required of how spin relaxation depends on the
system parameters and operating conditions. It is generally accepted
\cite{pikus:1984} that in bulk semiconductors with a zinc blende
structure and in quantum well (QW) structures based on these
materials the electronic spin relaxation is dominated by the
D'yakonov-Perel' \cite{dyakonov:1971, dyakonov:1972, pikus:1984,
dyakonov:1986, averkiev:2002, averkiev2:2002, averkiev:1999,
kainz:2003, lau:2001} (DP) and the Bir-Aronov-Pikus \cite{bir:1976}
mechanisms. In the present work, we consider $n$-doped
two-dimensional (2D) systems, where the DP mechanism becomes
dominant. This mechanism is intimately related to the spin splitting
of the electronic subbands in systems lacking inversion symmetry.
\cite{dyakonov:1986, averkiev:2002} The spin splitting can result
from the inversion asymmetry of the bulk crystal structure [bulk
inversion asymmetry (BIA)] \cite{dresselhaus:1955} and from the
inversion asymmetry of the QW structure [structure inversion
asymmetry (SIA)]. \cite{byc84} The interplay of BIA and SIA in
asymmetrically doped QWs gives rise to anisotropic spin relaxation
\cite{averkiev:2002, averkiev2:2002, averkiev:1999, kainz:2003} for
spins aligned along different crystallographic directions.

Previously, the DP theory has been formulated for the limiting cases
of low and high temperatures (i.e., for degenerate and nondegenerate
electron systems). \cite{dyakonov:1972, dyakonov:1986,
averkiev:2002, averkiev2:2002} Here we extend the theory to
arbitrary temperatures inbetween these limiting cases. Our
calculations show that often room temperature falls into this
intermediate regime. Therefore, our findings are particularly
important for spintronic devices working at room temperature.
\cite{rud03} We discuss the details of the temperature dependence
and address the influence of various momentum scattering mechanisms.
The temperature dependence of the spin relaxation is strongly
affected by the temperature dependence of the momentum scattering
mechanisms, which are ruled by different power laws in the electron
energy.

We present accurate calculations of the temperature dependent spin
relaxation rates assuming realistic system parameters. The spin
splittings are obtained from self-consistent calculations in the
multiband envelope-function approximation (EFA). \cite{winkler:1993}
We compare the calculated results with experimental data for
symmetric $n$-doped (001)-grown GaAs/AlGaAs QWs measured by
Terauchi \etal \cite{terauchi:1999} and Ohno \etal \cite{ohno:2000}
Theory and experiment agree very well.  Our calculations show that
the anisotropy of the spin relaxation persists up to room
temperature.

%%%%%%%%%%%%%%%%%%%%%%%%%%%%%%%%%%%%%%%%%%%%%%%%%%%%%%%%%%%%%%%%%%
\section{D'yakonov-Perel' mechanism and spin splitting}

In a single band approach, the DP spin relaxation is based on the
time evolution of the electron spin polarization
\begin{equation}
\vek{S} = \langle \trace (\varrho_{\vek{k}} \vek{\sigma}) \rangle =
\sum_{\vek{k}} \trace (\varrho_{\vek{k}} \vek{\sigma}),   
\end{equation}
where $\vek{k}$ is the electron wave vector, $\vek{\sigma}$ is the
vector of Pauli spin matrices, and $\varrho_{\vek{k}}$ is the $2 \times
2$ electron spin-density matrix. The dynamics of $\vek{S}$ is ruled
by a Bloch-like equation containing the spin-orbit coupling
\begin{equation}
  \label{socoupling}
  H_\mathrm{so}=\frac{\hbar}{2} \: \vek{\sigma}
  \cdot \vek{\Omega}(\vek{k}) \: .
\end{equation}
The spin-orbit coupling (\ref{socoupling}) is similar to a Zeeman
term, but $\vek{\Omega}(\vek{k})$ is an effective magnetic field
that depends on the underlying material, on the geometry of the device, 
and on the electron wave vector $\vek{k}$.  The vector $\trace
(\varrho_{\vek{k}} \vek{\sigma}) $ precesses about
$\vek{\Omega}(\vek{k})$ which results in spin relaxation. However,
(spin independent) momentum scattering of the electrons with, e.g.,
phonons, (nonmagnetic) impurities or other electrons
\cite{glazov:2002, brand:2002} changes $\vek{k}$ and thus the
direction and magnitude of $\vek{\Omega}(\vek{k})$ felt by the
electrons.  Therefore, frequent scattering (on the time scale of $|
\vek{\Omega}(\vek{k}) |^{-1}$) suppresses the precession and the
spin relaxation. This is the motional narrowing that is typical for
the DP mechanism, \cite{dyakonov:1972} according to which the spin
relaxation rate $\tau_s^{-1} \propto \tau_p$. Here $\tau_p$ is the
momentum scattering time.

Due to time reversal invariance we have $\vek{\Omega}(0) = 0$, while
for finite $\vek{k}$ the spin-orbit coupling $H_\mathrm{so}$ causes
a spin splitting $\hbar |\vek{\Omega}(\vek{k})|$ of the electron
states. In this article we focus on quasi 2D systems grown on a
$(001)$ surface and made from semiconductors with a zinc blende
structure. In leading order, the effective field in these systems
reads
\begin{equation}
\label{omega_formel}
\vek{\Omega}(\vek{k}_\|)
=
\frac{2\gamma}{\hbar} \left(\begin{array}{c}
    k_x \left(k_y^2 - \langle k_z^2 \rangle\right) \\[0.5ex]
    k_y \left(\langle k_z^2 \rangle - k_x^2\right) \\[0.5ex]
    0 
    \end{array} \right) +
\frac{2\alpha}{\hbar} \left(\begin{array}{c}
    k_y \\[0.5ex]
    -k_x \\[0.5ex]
    0
  \end{array} \right).
\end{equation}
The first term characterizes the BIA spin splitting of the electron
states. It is called the Dresselhaus or $k^3$ term.
\cite{dresselhaus:1955, braun:1985} It exists already in bulk
semiconductors with broken inversion symmetry. In quasi 2D systems
only the in-plane wave vector $\vek{k}_\| = (k_x, k_y, 0)$ is
a continuous variable.  In first order perturbation theory, the wave
vector components $k_z$ and powers thereof are replaced by
expectation values with respect to the subband wave functions.

In asymmetric QWs, SIA gives rise to the second term in Eq.\ 
(\ref{omega_formel}) which is frequently called the Rashba term.
\cite{byc84} The coefficients $\gamma$ and $\alpha$ depend on the
underlying semiconductor bulk material; but $\alpha$
depends also on the asymmetry of the QW in growth direction.

Within the DP mechanism the relaxation of the components of
$\vek{S}$ is ruled by \cite{dyakonov:1972}
\begin{equation}
\frac{\partial}{\partial t} S_i = - \frac{1}{\tau_i} S_i \, ,
\end{equation}  
where $i=z,+,-$ corresponds to the components of $\vek{S}$ along
$[001]$, $[110]$, and $[\overline{1}10]$, and we have $S_\pm =
1/\sqrt{2} ( S_x \pm S_y )$. The spin-relaxation rates are
\cite{averkiev2:2002}
\begin{subequations}
\label{sr_raten_gleichung}
\begin{align}
  \frac{1}{\tau_z} & = \frac{4}{\hbar^2} \hat{I} \tau_1
  \bigg[ k_\|^2 
  \left( \alpha^2 + \gamma^2 \langle k_z^2 \rangle^2 \right)
  - k_\|^4 \frac{\gamma^2 \langle k_z^2 \rangle}{2}
  \nonumber
  \\ & \qquad 
  +  k_\|^6 \gamma^2 \frac{1 + \tau_3/\tau_1}{16} \bigg]
  \; ,
  \\
  \frac{1}{\tau_\pm} & = \frac{2}{\hbar^2} \hat{I} \tau_1
  \bigg[ k_\|^2
  \left( \pm \alpha - \gamma \langle k_z^2 \rangle \right)^2
  + k_\|^4 \frac{\gamma}{2}
  \left( \pm \alpha - \gamma \langle k_z^2 \rangle \right) 
  \nonumber
  \\ & \qquad
  + k_\|^6 \gamma^2 \frac{1 + \tau_3/\tau_1}{16} \bigg]
  \; .
 \end{align}
\end{subequations}
The symbol $\hat{I}$ denotes an integral operator which is acting on
functions $f (\vek{k}_\|)$ according to
\begin{equation}
 \hat{I}f = \frac{1}{\int \left( F_+ - F_- \right) \dd^2 k_\|} 
 \int \left( F_+ - F_- \right) f(\vek{k}_\|) \: \dd^2 k_\|
 \; ,
 \label{integraloperator_gleichung}
\end{equation}
where 
\begin{equation}
F_\pm = \frac{1}{ 1+ \ee^{\beta[E(\vek{k}_\|) -\mu_\pm]} } \; ,
\end{equation}
is the Fermi distribution function, $\beta = 1 / (\kbolz T)$, and
$\mu_+$ and $\mu_-$ are the chemical potentials for up and down
spins, respectively. In Eq.\ (\ref{sr_raten_gleichung}), the
relaxation times $\tau_1$ and $\tau_3$ are given by
\begin{equation}
 \frac{1}{\tau_n(\vek{k}_\|)} = \sum_{\vek{k}'_\|} W (\vek{k}_\|, \vek{k}_\|')
 \left[ 1 - \cos (n \theta) \right]
 \; ,
 \label{eq:tau_2d}
\end{equation}
where $ W (\vek{k}_\|, \vek{k}_\|')$ is the momentum scattering rate
between the states with wave vectors $\vek{k}_\|$ and $\vek{k}'_\|$,
calculated according to Fermi's Golden Rule. \cite{lundstrom:2000}
The symbol $\theta$ denotes the angle between $\vek{k}_\|$ and
$\vek{k}'_\|$.

%%%%%%%%%%%%%%%%%%%%%%%%%%%%%%%%%%%%%%%%%%%%%%%%%%%%%%%%%%%%%%%%%%%
\section{D'yakonov-Perel' theory for arbitrary temperature}
\label{dp_beliebige_T_section}

In the following we use the effective mass approximation
\begin{equation}
  E (\mathbf{k}) = k^2 / \zeta \,,
 \label{effektivmassennaeherung_gleichung}
\end{equation}
where $\zeta = 2 m^\ast/\hbar^2$, so that we can switch from
$\vek{k}_\|$ to $E$ as integration variable in Eq.\ 
(\ref{sr_raten_gleichung}). Assuming a sufficiently small
\cite{delta_mu} difference $\Delta \mu = \mu_+ - \mu_-$, we have
$F_+ - F_- \approx - \Delta \mu \left(\partial F_0 / \partial
  E\right)$ and $\hat{I}$ can be simplified to
\begin{equation}
 \hat{I} f \approx \frac{1}{\int 
 \left( \frac{- \partial F_0}{\partial E}\right) \dd E} 
 \int \dd E \left(\frac{- \partial F_0}{\partial E}\right) f(E)
 \label{integraloperator_neu_gleichung}
\end{equation}
independent of $\Delta \mu$.
Here,
\begin{equation}
F_0 = \frac{1}{ 1+ \ee^{\beta[E -\mu_0]} }
\label{fermi_distribution_gleichung}
\end{equation}
is the equilibrium Fermi distribution with the temperature-dependent
chemical potential $\mu_0$. For a 2D system with parabolic energy
dispersion (\ref{effektivmassennaeherung_gleichung}) the chemical
potential $\mu_0$ can be exactly evaluated giving
\begin{equation}
 \label{eq:chempot}
 \mu_0 = \frac{1}{\beta} \ln\left(\ee^{E_\fermi \beta} -1 \right),
\end{equation}
where $E_\fermi$ is the Fermi energy at temperature $T=0$. 

To proceed further, we must specify the scattering times $\tau_1$
and $\tau_3$. Most scattering mechanisms \cite{lundstrom:2000,
singh:1993} have a power law dependence of $\tau_1$ and $\tau_3$ on
the energy $E$ with the same characteristic exponent $\nu$
\begin{subequations}
  \label{tau_von_E_gleichung}
\begin{eqnarray}
  \tau_1 & = & \Xi E^\nu \; ,
  \label{tau_1_von_E_gleichung}
  \\
  \tau_3 & \propto & E^\nu \; ,
\end{eqnarray}
\end{subequations}
where $\Xi$ is a constant independent of $E$. (Note, however, that
in general $\Xi$ depends on $T$.) Then, $\tau_3/\tau_1$ is a
constant describing the angular scattering characteristics of the
scattering mechanism (see
Appendix~\ref{winkelabhaengigkeit_abschnitt}).

A short calculation yields
\begin{equation}
 \hat{I} E^n =
 \frac{J_n(\beta \mu_0)}{\beta^n \left( 1 - \ee^{-E_\fermi \beta} \right) }
 \; ,
 \label{I_E_nu_gleichung}
\end{equation}
where
\begin{equation}
 J_n(z) = \int_0^\infty 
 \frac{x^n}{4 \cosh^2 \left( \frac{x-z}{2} \right) } \dd x
 \; .
 \label{J_nu_gleichung} 
\end{equation}
In particular, we find using Eq.\ (\ref{eq:chempot})
\begin{subequations}
  \begin{eqnarray}
 J_0(\beta \mu_0) & = & 1 - \ee^{-E_\fermi \beta} 
 \; ,
 \label{J_0_gleichung} 
 \\
 J_1(\beta \mu_0) & = & E_\fermi \beta
 \;.
 \label{J_1_gleichung} 
  \end{eqnarray}
\end{subequations}
For arbitrary $n$, the integrals $J_n (\beta \mu_0)$ can be
represented in closed form in terms of Lerch functions.
\cite{gradshteyn:1996} As this offers no advantage with respect to
the numerical evaluation, we calculate $J_n (\beta \mu_0)$ by
numerical quadrature.

The transport relaxation time $\tau_\mathrm{tr}$ can be obtained
from
\begin{align}
 \tau_\mathrm{tr} & = \frac{\displaystyle \int  E \tau_1(E) 
  \frac{\partial F_0}{\partial E} \dd E}
 {\displaystyle \int  E \frac{\partial F_0}{\partial E} \dd E}
 = \frac{ \hat{I} \Xi E^{\nu +1} }{\hat{I} E} 
 = \frac{\Xi  J_{\nu+1}(\beta \mu_0) }{\beta^{\nu+1} E_\fermi}
\; ,
\label{tau_transport_gleichung}
\end{align}
see Eq.\ (\ref{eq:transport_tau}) in the appendix. Using
\begin{equation}
 \hat{I} \tau_1 E^i 
 = \frac{\beta^{\nu+1} \tau_\mathrm{tr} E_\fermi \hat{I}
 E^{\nu+i}}{J_{\nu+1}(\beta \mu_0)}
 = \frac{\beta^{1-i} \tau_\mathrm{tr} E_\fermi}{1-\ee^{-E_\fermi \beta}} 
  \; \frac{ J_{\nu+i}(\beta \mu_0) }{J_{\nu+1}(\beta \mu_0)}
\end{equation}
where $i=1,2,3$, we simplify Eq.\ (\ref{sr_raten_gleichung}):
\begin{widetext}
\begin{subequations}
\begin{align}
 \frac{1}{\tau_z} & = \frac{4}{\hbar^2} 
 \frac{\beta E_\fermi}{1- \ee^{-E_\fermi \beta}} \tau_\mathrm{tr}
 \left[ 
 \left( \alpha^2 + \gamma^2 \langle k_z^2 \rangle^2 \right) 
 \frac{\zeta}{\beta}
 - \frac{\gamma^2 \langle k_z^2 \rangle}{2} 
   \left( \frac{\zeta}{\beta} \right)^2
   \frac{J_{\nu+2}(\beta \mu_0)}{J_{\nu+1}(\beta \mu_0)}
 + \gamma^2 \frac{1 + \tau_3/\tau_1}{16}  
   \left( \frac{\zeta}{\beta} \right)^3 
   \frac{J_{\nu+3}(\beta \mu_0)}{J_{\nu+1}(\beta \mu_0)}
 \right]
 \\
 \frac{1}{\tau_\pm} & = \frac{2}{\hbar^2}
 \frac{\beta E_\fermi}{1- \ee^{-E_\fermi \beta}} \tau_\mathrm{tr}
 \left[ 
 \left( \pm \alpha - \gamma \langle k_z^2 \rangle \right)^2  
 \frac{\zeta}{\beta}
 + \frac{\gamma}{2} 
   \left( \pm \alpha - \gamma \langle k_z^2 \rangle \right)
   \left( \frac{\zeta}{\beta} \right)^2  
   \frac{J_{\nu+2}(\beta \mu_0)}{J_{\nu+1}(\beta \mu_0)}
 + \gamma^2 \frac{1 + \tau_3/\tau_1}{16} 
   \left( \frac{\zeta}{\beta} \right)^3 
   \frac{J_{\nu+3}(\beta \mu_0)}{J_{\nu+1}(\beta \mu_0)}
 \right] .
 \end{align}
\label{spinrelaxation_verallgemeinert_gleichung}
\end{subequations}
\end{widetext}
These formulas generalize the hitherto \cite{averkiev2:2002} known
limiting cases of the low and high temperature regime of the DP spin
relaxation in quasi 2D systems.

%%%%%%%%%%%%%%%%%%%%%%%%%%%%%%%%%%%%%%%%%%%%%%%%%%%%%%%%%%%%%%%%%%
\section{Numerical Results}

\begin{figure*}[tb]
\includegraphics[width=\textwidth]{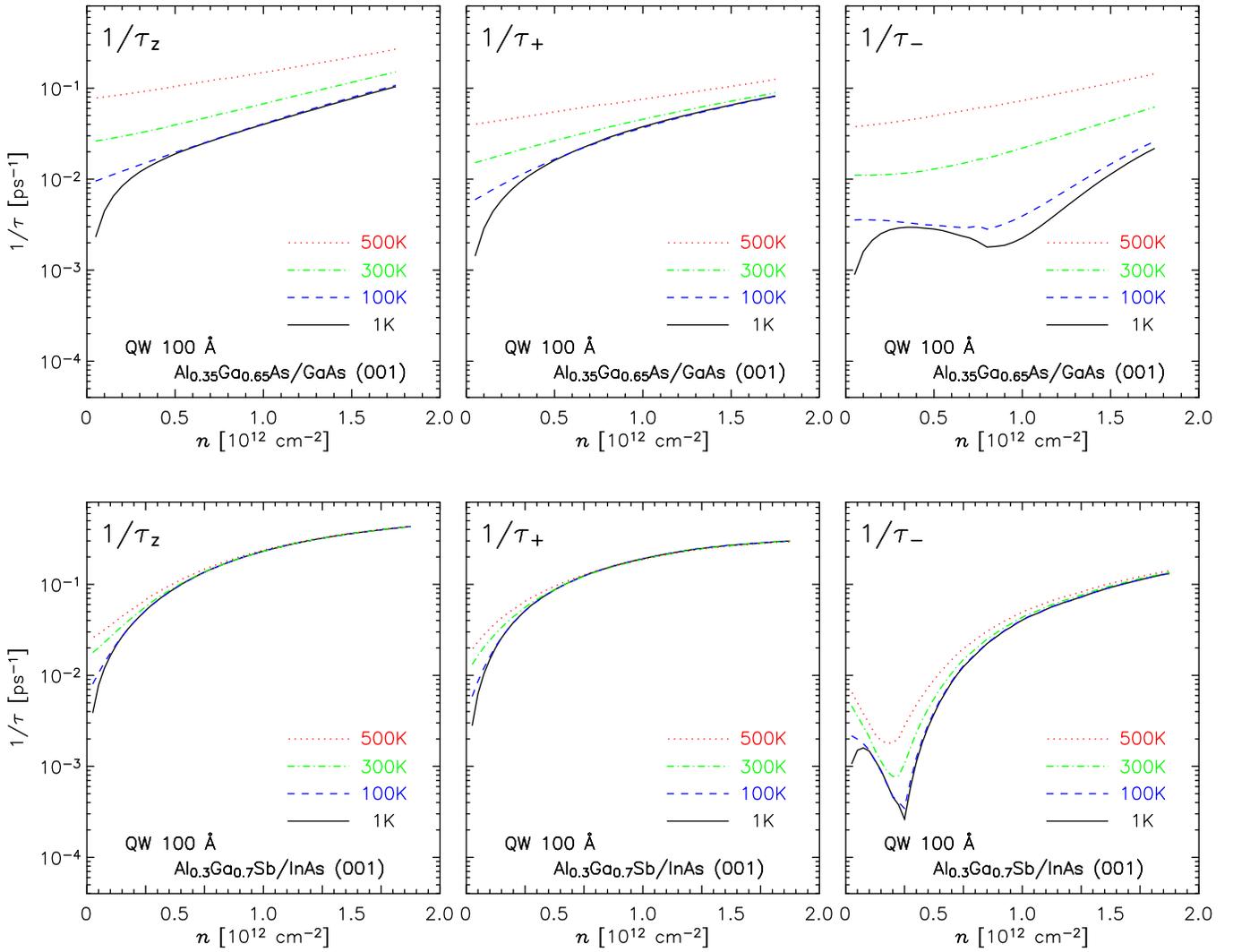}
\caption{
\label{spinrelax_variousT_abbildung}
Spin relaxation rates $\tau_z^{-1}$ and $\tau_\pm^{-1}$ as a
function of carrier density $n$ and for different temperatures $T$
in asymmetric (001) grown Al$_{0.35}$Ga$_{0.65}$As/GaAs-QWs (upper
row) and Al$_{0.3}$Ga$_{0.7}$Sb/InAs-QWs (lower row). The momentum
scattering was assumed to be isotropic (type~I, see
Table~\ref{streumechanismen_tabelle}) and independent of $T$. }
\end{figure*}

To illustrate the theory developed in the preceding section we show
in Fig.~\ref{spinrelax_variousT_abbildung} the spin relaxation rates
$\tau_z^{-1}$ and $\tau_\pm^{-1}$ calculated as a function of
carrier density $n$ and temperature $T$ for asymmetric (001) grown
AlGaAs/GaAs-QWs and AlGaSb/InAs-QWs with well width $L=100$~{\AA}.
The spin splitting was calculated self-consistently assuming that
the samples were $n$-doped on one side only. We used
$\tau_\mathrm{tr} = 0.1$~ps (Ref.~\onlinecite{rescale}) and we
assumed that the momentum scattering was isotropic. The parameters
are thus the same as in Ref.~\onlinecite{kainz:2003} so that our
results for $T=1$~K reproduce the values obtained in
Ref.~\onlinecite{kainz:2003} using the formulas for the low
temperature limit of DP spin relaxation. At moderately high
densities, the calculated rates usually increase with increasing
temperature by less than an order of magnitude. In
Fig.~\ref{spinrelax_variousT_abbildung} we have assumed that
$\tau_\mathrm{tr}$ does not depend on $T$. In real systems,
$\tau_\mathrm{tr}$ can decrease by more than an order of magnitude
for such an increase in temperature. Therefore, we can usually
expect smaller spin relaxation rates at room temperature as compared
to low temperatures.

The largest variation of the spin relaxation rates with temperature
occurs in the low density range. Here, the thermal energy $\kbolz T$
and the Fermi energy $E_\fermi$ are of the same order of magnitude
for low $T$. On the other hand, at higher temperatures the spin
relaxation is no longer controlled by $E_\fermi$ but by $\kbolz T$.
This results in the pronounced temperature dependence at low
densities which is visible in
Fig.~\ref{spinrelax_variousT_abbildung}. The weaker temperature
dependence obtained for the InAs-based systems in comparison with
GaAs-based systems can be related to the term proportional to $T^3$
in Eq.\ (\ref{spinrelaxation_verallgemeinert_gleichung}), which
dominates at low temperature for GaAs due to its larger effective
mass $m^\ast$ and smaller $\langle k_z^2 \rangle$.

The strong anisotropy of the spin relaxation times, i.e., the
difference between $\tau_+^{-1}$ and $\tau_-^{-1}$ and the
nonmonotonous behavior \cite{kainz:2003} persist up to temperatures
$T \sim 100$~K for the GaAs system and even far beyond room
temperature for the InAs system. The huge difference, previously
predicted for low temperature, \cite{kainz:2003} which is seen for
densities around $n \sim 0.6 \times 10^{12}$~cm$^{-2}$ in the
systems considered here, is therefore remarkably stable with
increasing temperature. For example, the rate $\tau_-^{-1}$ in the
InAs system at $n \sim 0.4 \times 10^{12}$~cm$^{-2}$ is
approximately fifty times smaller than $\tau_z^{-1}$ or
$\tau_-^{-1}$ even at room temperature.

%%%%%%%%%%%%%%%%%%%%%%%%%%%%%%%%%%%%%%%%%%%%%%%%%%%%%%%%%%%%%%%%%%
\section{Comparison with room temperature experiments}

\begin{figure}[tbp]
\includegraphics[width =\columnwidth]{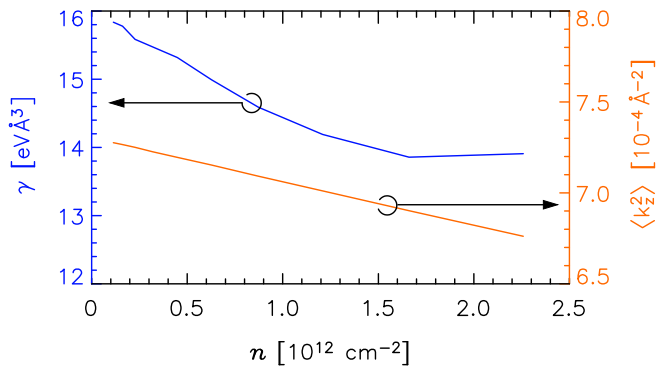}
\caption{
\label{spinaufspaltung_parameter_terauchi_abbildung}
Calculated BIA spin splitting parameters $\gamma$ (left axis) and
$\langle k_z^2 \rangle$ (right axis) as a function of carrier
density $n$ for a symmetrically doped Al$_{0.4}$Ga$_{0.6}$As/GaAs-QWs with
$L=75$~{\AA}.}
\end{figure}

Next, we calculate the spin relaxation times $\tau_z$ of
symmetrically $n$-doped (001)-grown QWs and compare our results with
the experimental data obtained by Terauchi \etal
\cite{terauchi:1999} and Ohno \etal \cite{ohno:2000} In both
experiments the authors used a pump-probe technique with circularly
polarized light. The Al$_{0.4}$Ga$_{0.6}$As/GaAs QW had a width
$L=75$~{\AA}. The sample parameters and measured \cite{tau_z_def}
$\tau_z$ for the eight samples investigated by Terauchi \etal\ are
reproduced in Table~\ref{terauchi_parameter_tabelle}.
For these systems we calculated the spin splitting and extracted the
values of the BIA spin-splitting parameters $\gamma$ and $\langle
k_z^2 \rangle$ as a function of the carrier density $n$ (see 
Fig.~\ref{spinaufspaltung_parameter_terauchi_abbildung}). The
details of this procedure have been described in Ref.\ 
\onlinecite{kainz:2003}.  Due to the symmetric doping of the
samples, there is no Rashba effect and $\alpha=0$. 

\begin{table}[bt]
\caption{
\label{terauchi_parameter_tabelle}
Measured carrier density $n$, Hall mobility $\mu_\mathrm{Hall}$ and
spin relaxation time $\tau_z$ of the samples investigated by
Terauchi \etal \cite{terauchi:1999} The samples (a), (b), and (c)
were $n$-doped in the barrier only, while (d), (e), and (f) were
$n$-doped in the well only. The samples (g) and (h) were $n$- and
(weakly) $p$-doped in the well. All values refer to room
temperature. }
 \extrarowheight 0.3ex
 \begin{tabular}{lcccc} 
  \hline \raisebox{3.0ex}{}%
  doping   & sample 
  & $n$ [$10^{11}$~cm$^{-2}$] 
  & $\mu_\mathrm{Hall}$ [cm$^2$/Vs] 
  & $\tau_z$ [ps]\rule[-2ex]{0pt}{0pt}
 \\ \hline \hline
  barrier      & (a) & 1.4 & 4600 & 33 \\
  $n$-doped    & (b) & 2.6 & 4700 & 32 \\
               & (c) & 4.7 & 4200 & 37 \\ \hline
  well         & (d) & 0.4 & 3500 & 35 \\ 
  $n$-doped    & (e) & 2.4 & 2800 & 43 \\
               & (f) & 6.0 & 2500 & 49 \\ \hline
well $n$- and  & (g) & 4.7 & 800 & 105 \\
$p$-doped      & (h) & 9.8 & 1500 & 56 \\ \hline \hline
 \end{tabular}
\end{table}

The spin relaxation rates in Eq.\ 
(\ref{spinrelaxation_verallgemeinert_gleichung}) depend on the
transport relaxation time $\tau_\mathrm{tr}$ which is related to the
transport mobility $\mu_\mathrm{tr}$ via
\begin{equation}
\mu_\mathrm{tr} = \frac{e \tau_\mathrm{tr}}{m^\ast} \: .
\label{transportbeweglichkeit_gleichung}
\end{equation}
Terauchi \etal \cite{terauchi:1999} determined for their samples the
Hall mobilities $\mu_\mathrm{Hall}$, see
Table~\ref{terauchi_parameter_tabelle}. These Hall mobilities are
equal to the transport mobilities only if the dominant scattering
mechanisms are isotropic [i.e., $W (\vek{k}_\|, \vek{k}_\|')$ does
not depend on the angle $\theta$]. If small-angle scattering
predominates the transport mobility $\mu_\mathrm{tr}$ and the Hall
mobility $\mu_\mathrm{Hall}$ differ by the Hall factor
$r_\mathrm{Hall}$. \cite{lundstrom:2000} A more detailed discussion
of the effects of various scattering mechanisms is given in
Appendix~\ref{streuung_anhang}. For the samples investigated by
Terauchi \etal \cite{terauchi:1999}, the dominant scattering
mechanisms were not known. Therefore, we have considered three
categories of scattering mechanisms listed in
Table~\ref{streumechanismen_tabelle} and denoted type I, II, and
III. Calculated values of the Hall factor $r_\mathrm{Hall}$ are
given in the same table.

Taking into account these effects and using the bulk GaAs effective
mass ($m^\ast/m_0 = 0.0665$), we obtain the transport relaxation
time $\tau_\mathrm{tr}$ from Eq.\ 
(\ref{transportbeweglichkeit_gleichung}). For the samples in
Table~\ref{terauchi_parameter_tabelle} the mobilities were measured
only at room temperature. In order to derive the results shown in
this section, we have thus used in Eq.\ 
(\ref{spinrelaxation_verallgemeinert_gleichung}) the same value of
$\tau_\mathrm{tr}$ for all temperatures. Accordingly, the
temperature dependence of the spin relaxation rates is determined
only by the Fermi distribution. A different experiment, where this
restriction does not apply, will be discussed below in
Sec.~\ref{tau_z_von_T_abschnitt}.

\begin{figure}[tbp]
\includegraphics[width=\columnwidth]{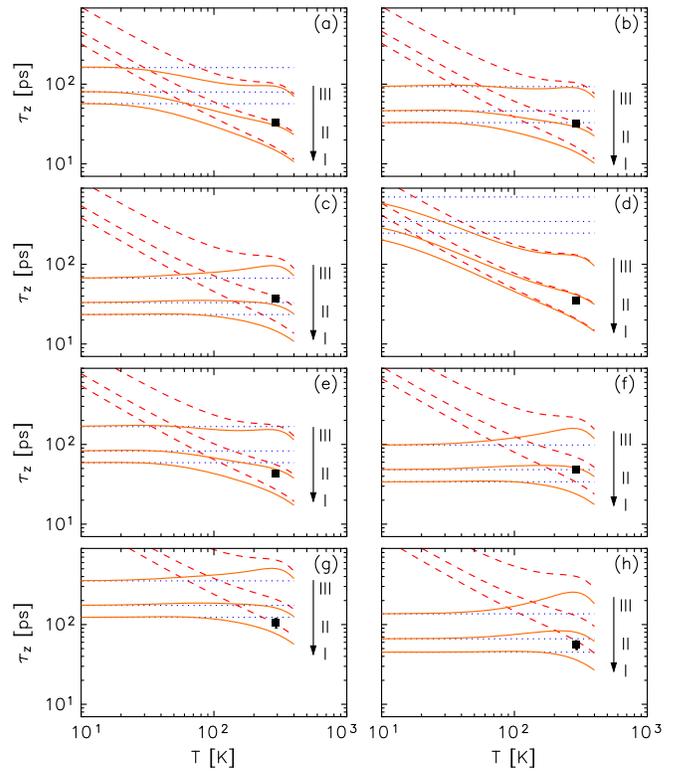}
\caption{
\label{spinrelax_terauchi_abbildung}
Calculated spin relaxation time $\tau_z$ of different samples
(a)--(h) as a function of temperature $T$ in comparison with room
temperature measurements (squares) from Terauchi \etal
\cite{terauchi:1999} The theoretical results refer to the degenerate
limit (dotted lines), to the nondegenerate (dashed lines) limit and
to the generalized theory (full lines).  For each set of curves, the
upper curve corresponds to scattering at weakly screened ionized
impurities (type~III), while the middle and lower curves represent
momentum scattering of type II and I, respectively (see
Tab.~\ref{streumechanismen_tabelle}).}
\end{figure}

The calculated results are shown in
Fig.~\ref{spinrelax_terauchi_abbildung} together with the
experimental data of Terauchi \etal\ For comparison, we show also
the results based on the formulas for the low and high temperature
limit, respectively. It can be seen that Eq.\ 
(\ref{spinrelaxation_verallgemeinert_gleichung}) reproduces the
results obtained with the simplified formulas for the corresponding
limits. For most of the systems investigated by Terauchi \etal, room
temperature falls into the nondegenerate (high temperature) limit.
However, in the more heavily doped samples a significant deviation
appears. For sample (h) the value of $\tau_z$ predicted by the
theory for the nondegenerate limit is approximately $60$\% larger
than the value obtained using Eq.\ 
(\ref{spinrelaxation_verallgemeinert_gleichung}).

In GaAs at room temperature, the mobility is limited by scattering
at the deformation potential of acoustic phonons (type~I) and at
polar optical phonons (type~II). The scattering at ionized
impurities (weakly screened: type~II; screened: type~I), which
usually dominates the mobility at low $T$, is not relevant here.
Therefore, the results calculated for scattering mechanisms of
type~III do not apply in the present case.

The experimental values for the spin relaxation time are between the
theoretical results obtained for scattering mechanisms of type~I and
II. For the samples (g) and (h) the measured values are
significantly closer to the theoretical values obtained for type~I
scattering than for all other samples. This agrees well with the
fact that only these two samples were codoped with donors and
acceptors. The scattering at neutral acceptors makes type~I
scattering more important for these samples. This reasoning is
supported by the fact that the samples (g) and (h) have the smallest
mobilities (see Tab. \ref{streumechanismen_tabelle}) which we
attribute to the additional scattering mechanism.

In summary, we have achieved quantitative agreement between theory
and experiment for the spin relaxation rates of eight different
samples. A similar EFA-based approach was applied by Lau \etal
\cite{lau:2001} for the samples discussed here with comparable
results. However, the previous work did not take into account the
doping self-consistently and no explicit formulas were given for the
treatment of the temperature dependence of the spin relaxation.
Furthermore, we would like to emphasize that it is important to take
into account the difference between the Hall mobility usually
measured in experiments and the transport mobility, which governs
the spin relaxation. To the best of our knowledge, so far this
aspect has not been taken into account in the context of spin
relaxation.

%%%%%%%%%%%%%%%%%%%%%%%%%%%%%%%%%%%%%%%%%%%%%%%%%%%%%%%%%%%%%%%%%%
\section{Comparison with temperature dependent measurements}
\label{tau_z_von_T_abschnitt}

\begin{figure}[tb]
\includegraphics[width=\columnwidth]{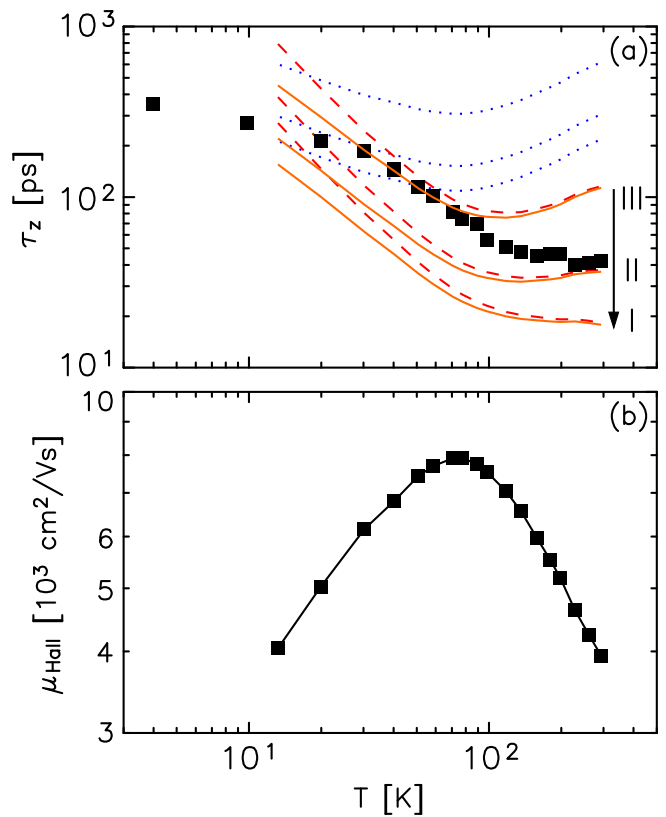}
\caption{
\label{spinrelax_ohno_abbildung}
(a) Spin relaxation time $\tau_z$ and (b) measured \cite{ohno:2000,
ohno:2003} Hall mobility $\mu_\mathrm{Hall}$ as a function of
temperature $T$. In (a) the labeling of the calculated curves is
analogous to Fig.~\ref{spinrelax_terauchi_abbildung} and the
measured values \cite{ohno:2000} are represented by squares. The
sample parameters are those of sample (d) in
Table~\ref{terauchi_parameter_tabelle}. In the very low temperature
range $T < 13$~K, no theoretical evaluation of $\tau_z$ is possible
due to lack of experimental data for the Hall mobility. }
\end{figure}

The spin relaxation time depends on the temperature via the Fermi
distribution function and the variation of the transport mobility.
To get a more detailed understanding of the interplay of these
quantities we compare our theoretical results with the spin
relaxation times measured as a function of temperature by Ohno \etal
\cite{ohno:2000} The sample used in their experiment corresponds to
sample (d) of Table~\ref{terauchi_parameter_tabelle}.
The measured values of the spin relaxation time
$\tau_z$ and the Hall mobility $\mu_\mathrm{Hall}$ are reproduced in
Fig.\ \ref{spinrelax_ohno_abbildung}. Once again, the spin
relaxation time $\tau_z$ has been calculated as a function of
temperature. But in contrast to
Fig.~\ref{spinrelax_terauchi_abbildung}, we have used here the
temperature dependent transport relaxation time obtained from the
Hall mobility measured as a function of~$T$. Figure
\ref{spinrelax_ohno_abbildung}(a) shows the theoretical results of
the extended theory for the three types of scattering mechanisms
together with the results assuming the degenerate and nondegenerate
limit.

Once again, an excellent quantitative agreement between theory and
experiment can be achieved at room temperature assuming that the
dominant scattering mechanisms are of type~II (e.g., polar optical
phonons). The measured decrease of the Hall mobility at high
temperatures is a signature of phonon scattering,
\cite{lundstrom:2000} which agrees well with this interpretation.

In the low temperature range (up to about $100$~K) the experimental
data are well explained by type~III scattering (i.e., by scattering
at weakly screened impurities). In samples which are not modulation
doped, like the ones considered here, this scattering mechanism
usually limits the mobility at low temperatures and leads to a
characteristic increase of the mobility with increasing temperature
[Fig.~\ref{spinrelax_ohno_abbildung}(b)].

The temperature dependence of the Hall mobility at intermediate
temperatures ($80$~K${}\lesssim T \lesssim 200$~K) indicates that
ionized impurity scattering becomes less important with increasing
temperature. Instead, phonon scattering becomes the dominant
scattering mechanism. As expected, the measured values of
$\tau_z(T)$ depart from the calculated curve for type~III scattering
with increasing temperature and approach the curve for type~II
scattering (polar phonon scattering). While we have a remarkably
good agreement between theory and experiment for $T \gtrsim 20$~K, a
discrepancy arises at low temperatures $T \lesssim 20$~K. As the
Hall mobility was not measured for $T<13$~K this deviation cannot be
examined more closely. It could be caused by approximations in the
scattering theory. For example, at low temperatures the simple
power-law dependence in Eq.\ (\ref{tau_von_E_gleichung}) becomes
incorrect for ionized impurity scattering. \cite{lundstrom:2000} In
samples with high mobilities (mobilities larger than the mobilities
of the samples studied in this work) the electron-electron
collisions neglected here constitute a further possible source of
systematic errors. Such processes conserve the total momentum of the
electron system so that they do not affect the mobility. The DP spin
relaxation, on the other hand, is reduced by any kind of scattering.
We therefore have a correction due to electron-electron scattering,
which is particularly important in high mobility samples.
\cite{glazov:2003, glazov:2002, glazov:cond-mat:2003} Finally, it is
possible that in the pump-probe experiment by Ohno \etal
\cite{ohno:2000} the laser beam heated up the electron system so
that the electron temperature was indeed higher than shown in
Fig.~\ref{spinrelax_ohno_abbildung}(a). \cite{oest04}

%%%%%%%%%%%%%%%%%%%%%%%%%%%%%%%%%%%%%%%%%%%%%%%%%%%%%%%%%%%%%%%%%%
\section{Conclusions}

Based on the self-consistent multiband envelope function
approximation, we obtain an accurate description of the spin
splitting and D'yakonov-Perel' spin relaxation in quasi 2D systems.
In the present work, we extend the theory of DP spin relaxation in
order to connect the previously known results for the limiting cases
of low and high temperatures. Therefore, our realistic theory
renders considerations with respect to the degeneracy or
nondegeneracy of the electron system unnecessary.

We demonstrate the strengths of our approach by comparing with spin
relaxation times $\tau_z$ measured by Terauchi \etal
\cite{terauchi:1999} and Ohno \etal \cite{ohno:2000} for a large
variety of different samples. Using a simple model,
\cite{dyakonov:1986} the latter authors obtained theoretical
estimates for $\tau_z$ that differed by an order of magnitude from
the experimental values, whereas we are able to obtain excellent
quantitative agreement between experiment and theory.

Up to now, the spin relaxation times $\tau_\pm$ could not be
determined experimentally. We corroborate here the prediction
\cite{averkiev:1999, kainz:2003} that one of the times $\tau_+$ or
$\tau_-$ increases nonmonotonously as a function of electron
density. Our temperature-dependent calculations show that this
feature is remarkably stable with increasing temperature and that it
is connected with a huge anisotropy of the spin relaxation.

%%%%%%%%%%%%%%%%%%%%%%%%%%%%%%%%%%%%%%%%%%%%%%%%%%%%%%%%%%%%%%%%%%
\begin{acknowledgments}
  We would like to thank L.~E.~Golub for valuable discussions.
  R.~W.\ also appreciates stimulating discussions with M.~Oestreich
  and D.~H\"agele. This work has been supported by the DFG via
  Forschergruppe~370 {\em Ferromagnet-Halbleiter-Nanostrukturen} and
  by BMBF.
\end{acknowledgments}

%%%%%%%%%%%%%%%%%%%%%%%%%%%%%%%%%%%%%%%%%%%%%%%%%%%%%%%%%%%%%%%%%%
\appendix
\section{Scattering theory}
\label{streuung_anhang}

In this appendix we give a short overview of the energy dependence
and scattering angle characteristic of various electron momentum
scattering mechanisms. As commonly done in the literature,
\cite{lundstrom:2000, singh:1993, gantmakher:1987, ridley:1993,
ferreira:1989} we assume that the electron system is nondegenerate
and that the scattering is quasi-elastic, i.e., the scattering does
not change the kinetic energy. We start by considering bulk systems.
Then we discuss the modifications necessary for 2D systems.

%%%%%%%%%%%%%%%%%%%%%%%%%%%%%%%%%%%%%%%%%%%%%%%%%%%%%%%%%%%%%%%%%%
\subsection{Relaxation times}

In this work we treat scattering based on Fermi's Golden Rule, where
the transition rate $W (\vek{k},\vek{k}')$ between states with wave
vectors $\vek{k}$ and $\vek{k}'$ is \cite{lundstrom:2000}
\begin{subequations}
\begin{align}
W (\vek{k},\vek{k}')  = & \frac{2 \pi}{\hbar} 
 \left| H^{\mathrm{scatt}}_{\vek{k}' \vek{k}} \right|^2
 \delta \left[ E(\vek{k}') - E(\vek{k}) \right]
 \,, 
 \\
H^{\mathrm{scatt}}_{\vek{k}' \vek{k}}  = & \frac{1}{\Omega} 
 \int_{\Omega} \ee^{-i \vek{k}'\cdot\vek{r}} \, 
 U^{\mathrm{scatt}}(\vek{r}) \,  \ee^{i \vek{k}\cdot\vek{r}}
 \dd^3 r
\end{align}
\end{subequations}
with the crystal volume $\Omega$ and the scattering potential
$U^{\mathrm{scatt}}(\vek{r})$. Using the effective mass
approximation (\ref{effektivmassennaeherung_gleichung}), we obtain
for the scattering rate out of the state $|\vek{k} \rangle$ into any
other state $|\vek{k}' \rangle$ of a nondegenerate electron system
\cite{lundstrom:2000}
\begin{equation}
 \label{eq:scatt_rate}
 \frac{1}{\tau(\vek{k})} = \sum_{\vek{k}'}^\sim W (\vek{k}, \vek{k}') 
 \; .
\end{equation}
Here the tilde indicates that the sum is restricted to final states
$|\vek{k}' \rangle$ with the same spin orientation as the initial
state, i.e., we have assumed that the scattering processes do not
change the spin. The scattering time $\tau(\vek{k})$ is the average
time between scattering events, i.e., it can be identified with the
time during which the electron is in the state~$|\vek{k}\rangle$.

For non-isotropic scattering mechanisms the information about the
initial momentum is not lost after the time $\tau(\vek{k})$, but
decays on the time scale of the momentum relaxation time
$\tau_1(\vek{k})$, which is defined as \cite{lundstrom:2000}
\begin{equation}
 \frac{1}{\tau_1(\vek{k})} = \sum_{\vek{k}'}^\sim W (\vek{k}, \vek{k}') 
 \left( 1 - \cos{\theta} \right)
\; .
\label{tau1_gleichung}
\end{equation}
For elastic scattering, as assumed here, $\tau_1(\vek{k})$ depends
only on the energy $E(\vek{k})$ (see
Sec.~\ref{winkelabhaengigkeit_abschnitt}). The time $\tau_1$
determines the transport mobility via Eq.\ 
(\ref{transportbeweglichkeit_gleichung}) and \cite{averkiev2:2002,
lundstrom:2000}
\begin{equation}
  \label{eq:transport_tau}
  \tau_\mathrm{tr} =
  \frac{\displaystyle\int \dd E Z(E) E 
        \frac{\partial F_0}{\partial E} \tau_1(E)}
       {\displaystyle\int \dd E Z(E) E 
        \frac{\partial F_0}{\partial E} }
  \; .
\end{equation}
Here, $F_0$ is the Fermi distribution function
(\ref{fermi_distribution_gleichung}). Assuming a parabolic
dispersion (\ref{effektivmassennaeherung_gleichung}), the density of
states $Z(E)$ is proportional to $\sqrt{E}$ in 3D systems and it is
independent of $E$ in 2D systems.

%%%%%%%%%%%%%%%%%%%%%%%%%%%%%%%%%%%%%%%%%%%%%%%%%%%%%%%%%%%%%%%%%%
\subsection{Scattering angle characteristic}
\label{winkelabhaengigkeit_abschnitt}

In the framework of the aforementioned approximations, the
transition rate $W (\vek{k}, \vek{k}')$ can be expressed for many
scattering mechanisms as a power law \cite{gantmakher:1987}
\begin{equation}
  \label{eq:rate_q}
  W (\vek{k}, \vek{k}') \propto \frac{1}{q^{2\nu}} \; ,
\end{equation}
where $q$ is the momentum transfer
\begin{subequations}
  \label{eq:impulsuebertrag}
  \begin{align}
    q \equiv |\vek{k} - \vek{k}'| & = 2 k \sin (\theta / 2) \\
    & = 2 \sqrt {\zeta \, E} \, \sin (\theta / 2) \; .
  \end{align}
\end{subequations}
Combining these equations with Eq.\ (\ref{eq:tau_2d}) we get Eq.\ 
(\ref{tau_von_E_gleichung}). In Table~\ref{streumechanismen_tabelle}
we have compiled from the literature \cite{singh:1993,
gantmakher:1987, lundstrom:2000, ridley:1993} the exponents $\nu$
for various scattering mechanisms. The scattering mechanisms of
type~I are isotropic, while for type II and III small angle
scattering is increasingly predominant. The values of $\nu$ for 2D
systems are derived from the values for 3D systems, using the fact
that the momentum scattering rates $1 / \tau_1$ are proportional to
the density of states. \cite{lundstrom:2000} Assuming that the
remaining contributions to the exponent $\nu$ do not depend on the
dimensionality of the system, we thus have
\begin{equation}
\nu^\mathrm{2D}=\nu^\mathrm{3D}+1/2 \: .
\label{s2D_gleichung}
\end{equation}
For some scattering mechanisms, values of $\nu^\mathrm{2D}$ have
been given explicitly in the literature. \cite{hess:1979, lee:1983}
They agree with our values in
Table~\ref{streumechanismen_tabelle}.

\begin{table*}[tbp]
\caption{\label{streumechanismen_tabelle}
Energy dependence, scattering angle characteristic, and Hall factor
for various scattering mechanisms. We have listed the exponents for
the power law behavior of the momentum relaxation time [Eq.\ 
(\ref{tau_von_E_gleichung})] for bulk \cite{singh:1993,
gantmakher:1987, lundstrom:2000, ridley:1993} ($\nu^\mathrm{3D}$)
and 2D systems ($\nu^\mathrm{2D}$) based on Eq.\ (\ref{s2D_gleichung}). The 
scattering angle characteristic is specified by the ratio of the
scattering times $\tau_3/\tau_1$ (cf.\ 
Sect.~\ref{winkelabhaengigkeit_abschnitt}). The Hall factor
$r_\mathrm{Hall}$ is the ratio of Hall and transport mobility
(cf.~Sect.~\ref{beweglichkeit_abschnitt}). }
 \tabcolsep 1em
 \renewcommand{\arraystretch}{1.4}
 \begin{tabular}{cm{6cm}*{5}{>{$}c<{$}}}
  \hline
  & & \multicolumn{2}{c}{3D} &\multicolumn{3}{c}{2D}
 \\
  & scattering mechanisms
  & \nu^\mathrm{3D}
  & r_\mathrm{Hall}^\mathrm{3D}
  & \nu^\mathrm{2D}
  & \tau_3/\tau_1
  & r_\mathrm{Hall}^\mathrm{2D}
  \\ \hline \hline
type~I: & 
 acoustic phonons (deformation potential) \newline
 optical phonons (deformation potential) \newline
 ionized impurities (screened) \newline
 neutral impurities \newline
 alloy scattering \newline
 interface roughness
&  -\frac{1}{2}       
& \frac{3\pi}{8} & 0 & 1 & 1\\ \hline
type~II: &
 acoustic phonons (polar, piezoelectric) \newline
 optical phonons (polar)
&  +\frac{1}{2}       
& \frac{45 \pi}{128} & 1 & \frac{1}{3} & \frac{7}{5}\\ \hline
type~III: & 
 ionized impurities (weakly screened)
&  +\frac{3}{2}      
& \frac{415 \pi}{512} & 2 & \frac{1}{9} & \frac{99}{35} \\ \hline
  \hline
 \end{tabular}
\end{table*}

Using Eqs.\ (\ref{eq:tau_2d}), (\ref{eq:rate_q}), and
(\ref{eq:impulsuebertrag}) we have in 2D systems
\begin{equation}
  \label{eq:tau_n_explicit}
  \frac{1}{\tau_n} \propto \int_0^{2\pi} \dd \theta \,
  \frac{1 - \cos (n\theta)}{\sin^{2\nu} (\theta / 2)} \;.
\end{equation}
Thus we get \cite{pikus:1995}
\begin{equation}
  \label{eq:tau_ratio}
  \frac{\tau_3}{\tau_1} = \left\{
    \begin{array}{l@{\hspace{2em}}l}
      \displaystyle
      \frac{(2 - \nu)(3 - \nu)}{\nu^2 - \nu + 6} & \nu \le 3/2\\[2ex]
      1/9 & \nu > 3/2 \;.
    \end{array}\right.
\end{equation}
The values for the ratio $\tau_3/\tau_1$ obtained by means of Eq.\ 
(\ref{eq:tau_ratio}) are listed in
Table~\ref{streumechanismen_tabelle}. They are identical with the
values used by Lau \etal \cite{lau:2001} as far as they are given
there.

%%%%%%%%%%%%%%%%%%%%%%%%%%%%%%%%%%%%%%%%%%%%%%%%%%%%%%%%%%%%%%%%%%
\subsection{Hall and transport mobility}
\label{beweglichkeit_abschnitt}

In general, the mobility $\mu_\mathrm{Hall}$ determined from Hall
measurements differs from the transport mobility $\mu_\mathrm{tr}$
(obtained from, e.g., the experimental conductance) by the Hall
factor \cite{singh:1993, lundstrom:2000}
\begin{equation}
r_\mathrm{Hall} = \frac{\mu_\mathrm{Hall}}{\mu_\mathrm{tr}}
\; .
\end{equation}
The values of $r_\mathrm{Hall}$ in
Table~\ref{streumechanismen_tabelle} have been obtained from the
exponent $\nu$ (for 3D and 2D systems, respectively) using
\cite{singh:1993, lundstrom:2000}
\begin{equation}
r_\mathrm{Hall} = \frac{ \Gamma(2 \nu + 5/2) \Gamma(5/2)}
                       {\left[ \Gamma(\nu + 5/2) \right]^2} \;,
\end{equation}
where $\Gamma$ denotes the Gamma function. \cite{gradshteyn:1996} We
have $1 \le r_\mathrm{Hall} \le 99/35$, i.e., $\mu_\mathrm{tr} \le
\mu_\mathrm{Hall}$.

%%%%%%%%%%%%%%%%%%%%%%%%%%%%%%%%%%%%%%%%%%%%%%%%%%%%%%%%%%%%%%%%%%


\begin{thebibliography}{40}
\expandafter\ifx\csname natexlab\endcsname\relax\def\natexlab#1{#1}\fi
\expandafter\ifx\csname bibnamefont\endcsname\relax
  \def\bibnamefont#1{#1}\fi
\expandafter\ifx\csname bibfnamefont\endcsname\relax
  \def\bibfnamefont#1{#1}\fi
\expandafter\ifx\csname citenamefont\endcsname\relax
  \def\citenamefont#1{#1}\fi
\expandafter\ifx\csname url\endcsname\relax
  \def\url#1{\texttt{#1}}\fi
\expandafter\ifx\csname urlprefix\endcsname\relax\def\urlprefix{URL }\fi
\providecommand{\bibinfo}[2]{#2}
\providecommand{\eprint}[2][]{\url{#2}}

\bibitem[{\citenamefont{DiVincenzo}(1995)}]{divincenzo:1995}
\bibinfo{author}{\bibfnamefont{D.~P.} \bibnamefont{DiVincenzo}},
  \bibinfo{journal}{Science} \textbf{\bibinfo{volume}{270}},
  \bibinfo{pages}{255} (\bibinfo{year}{1995}).

\bibitem[{\citenamefont{Prinz}(1998)}]{prinz:1998}
\bibinfo{author}{\bibfnamefont{G.~A.} \bibnamefont{Prinz}},
  \bibinfo{journal}{Science} \textbf{\bibinfo{volume}{282}},
  \bibinfo{pages}{1660} (\bibinfo{year}{1998}).

\bibitem[{\citenamefont{Wolf et~al.}(2001)\citenamefont{Wolf, Awschalom,
  Buhrman, Daughton, {von Moln\'ar}, Roukes, Chtchelkanova, and
  Treger}}]{wol01}
\bibinfo{author}{\bibfnamefont{S.~A.} \bibnamefont{Wolf}},
  \bibinfo{author}{\bibfnamefont{D.~D.} \bibnamefont{Awschalom}},
  \bibinfo{author}{\bibfnamefont{R.~A.} \bibnamefont{Buhrman}},
  \bibinfo{author}{\bibfnamefont{J.~M.} \bibnamefont{Daughton}},
  \bibinfo{author}{\bibfnamefont{S.}~\bibnamefont{{von Moln\'ar}}},
  \bibinfo{author}{\bibfnamefont{M.~L.} \bibnamefont{Roukes}},
  \bibinfo{author}{\bibfnamefont{A.~Y.} \bibnamefont{Chtchelkanova}},
  \bibnamefont{and} \bibinfo{author}{\bibfnamefont{D.~M.}
  \bibnamefont{Treger}}, \bibinfo{journal}{Science}
  \textbf{\bibinfo{volume}{294}}, \bibinfo{pages}{1488} (\bibinfo{year}{2001}).

\bibitem[{\citenamefont{Awschalom et~al.}(2002)\citenamefont{Awschalom, Loss,
  and Samarth}}]{aws02}
\bibinfo{editor}{\bibfnamefont{D.~D.} \bibnamefont{Awschalom}},
  \bibinfo{editor}{\bibfnamefont{D.}~\bibnamefont{Loss}}, \bibnamefont{and}
  \bibinfo{editor}{\bibfnamefont{N.}~\bibnamefont{Samarth}}, eds.,
  \emph{\bibinfo{title}{Semiconductor Spintronics and Quantum Computations}}
  (\bibinfo{publisher}{Springer}, \bibinfo{address}{Berlin},
  \bibinfo{year}{2002}).

\bibitem[{\citenamefont{Oestreich et~al.}(2002)\citenamefont{Oestreich, Bender,
  H\"ubner, H\"agele, R\"uhle, Hartmann, Klar, Heimbrodt, Lampalzer, Volz
  et~al.}}]{oes02}
\bibinfo{author}{\bibfnamefont{M.}~\bibnamefont{Oestreich}},
  \bibinfo{author}{\bibfnamefont{M.}~\bibnamefont{Bender}},
  \bibinfo{author}{\bibfnamefont{J.}~\bibnamefont{H\"ubner}},
  \bibinfo{author}{\bibfnamefont{D.}~\bibnamefont{H\"agele}},
  \bibinfo{author}{\bibfnamefont{W.~W.} \bibnamefont{R\"uhle}},
  \bibinfo{author}{\bibfnamefont{T.}~\bibnamefont{Hartmann}},
  \bibinfo{author}{\bibfnamefont{P.~J.} \bibnamefont{Klar}},
  \bibinfo{author}{\bibfnamefont{W.}~\bibnamefont{Heimbrodt}},
  \bibinfo{author}{\bibfnamefont{M.}~\bibnamefont{Lampalzer}},
  \bibinfo{author}{\bibfnamefont{K.}~\bibnamefont{Volz}}, \bibnamefont{et~al.},
  \bibinfo{journal}{Semicond.\ Sci.\ Technol.} \textbf{\bibinfo{volume}{17}},
  \bibinfo{pages}{285} (\bibinfo{year}{2002}).

\bibitem[{\citenamefont{Pikus and Titkov}(1984)}]{pikus:1984}
\bibinfo{author}{\bibfnamefont{G.~E.} \bibnamefont{Pikus}} \bibnamefont{and}
  \bibinfo{author}{\bibfnamefont{A.~N.} \bibnamefont{Titkov}}, in
  \emph{\bibinfo{booktitle}{Optical Orientation}}, edited by
  \bibinfo{editor}{\bibfnamefont{F.}~\bibnamefont{Meier}} \bibnamefont{and}
  \bibinfo{editor}{\bibfnamefont{B.~P.} \bibnamefont{Zakharchenya}}
  (\bibinfo{publisher}{Elsevier}, \bibinfo{address}{Amsterdam},
  \bibinfo{year}{1984}), chap.~\bibinfo{chapter}{3}, pp.
  \bibinfo{pages}{73--131}.

\bibitem[{\citenamefont{D'yakonov and Perel'}(1971)}]{dyakonov:1971}
\bibinfo{author}{\bibfnamefont{M.~I.} \bibnamefont{D'yakonov}}
  \bibnamefont{and} \bibinfo{author}{\bibfnamefont{V.~I.}
  \bibnamefont{Perel'}}, \bibinfo{journal}{Sov. Phys. JETP}
  \textbf{\bibinfo{volume}{33}}, \bibinfo{pages}{1053} (\bibinfo{year}{1971}),
  \bibinfo{note}{[Zh. Eksp. Teor. Fiz. {\bf 60}, 1954 (1971)]}.

\bibitem[{\citenamefont{D'yakonov and Perel'}(1972)}]{dyakonov:1972}
\bibinfo{author}{\bibfnamefont{M.~I.} \bibnamefont{D'yakonov}}
  \bibnamefont{and} \bibinfo{author}{\bibfnamefont{V.~I.}
  \bibnamefont{Perel'}}, \bibinfo{journal}{Sov. Phys. Solid State}
  \textbf{\bibinfo{volume}{13}}, \bibinfo{pages}{3023} (\bibinfo{year}{1972}),
  \bibinfo{note}{[Fiz. Tverd. Tela, {\bf 13}, 3581 (1971)]}.

\bibitem[{\citenamefont{D'yakonov and Kachorovskii}(1986)}]{dyakonov:1986}
\bibinfo{author}{\bibfnamefont{M.~I.} \bibnamefont{D'yakonov}}
  \bibnamefont{and} \bibinfo{author}{\bibfnamefont{V.~Y.}
  \bibnamefont{Kachorovskii}}, \bibinfo{journal}{Sov. Phys. Semicond.}
  \textbf{\bibinfo{volume}{20}}, \bibinfo{pages}{110} (\bibinfo{year}{1986}),
  \bibinfo{note}{[Fiz. Tekh. Poluprovodn. {\bf 20}, 178 (1986)]}.

\bibitem[{\citenamefont{Averkiev
  et~al.}(2002{\natexlab{a}})\citenamefont{Averkiev, Golub, and
  Willander}}]{averkiev:2002}
\bibinfo{author}{\bibfnamefont{N.~S.} \bibnamefont{Averkiev}},
  \bibinfo{author}{\bibfnamefont{L.~E.} \bibnamefont{Golub}}, \bibnamefont{and}
  \bibinfo{author}{\bibfnamefont{M.}~\bibnamefont{Willander}},
  \bibinfo{journal}{J.\ Phys.: Condens.\ Matter} \textbf{\bibinfo{volume}{14}},
  \bibinfo{pages}{R271} (\bibinfo{year}{2002}{\natexlab{a}}).

\bibitem[{\citenamefont{Averkiev and Golub}(1999)}]{averkiev:1999}
\bibinfo{author}{\bibfnamefont{N.~S.} \bibnamefont{Averkiev}} \bibnamefont{and}
  \bibinfo{author}{\bibfnamefont{L.~E.} \bibnamefont{Golub}},
  \bibinfo{journal}{Phys.\ Rev.~B} \textbf{\bibinfo{volume}{60}},
  \bibinfo{pages}{15582} (\bibinfo{year}{1999}).

\bibitem[{\citenamefont{Kainz et~al.}(2003)\citenamefont{Kainz, R\"ossler, and
  Winkler}}]{kainz:2003}
\bibinfo{author}{\bibfnamefont{J.}~\bibnamefont{Kainz}},
  \bibinfo{author}{\bibfnamefont{U.}~\bibnamefont{R\"ossler}},
  \bibnamefont{and} \bibinfo{author}{\bibfnamefont{R.}~\bibnamefont{Winkler}},
  \bibinfo{journal}{Phys.\ Rev.~B} \textbf{\bibinfo{volume}{68}},
  \bibinfo{pages}{075322} (\bibinfo{year}{2003}).

\bibitem[{\citenamefont{Averkiev
  et~al.}(2002{\natexlab{b}})\citenamefont{Averkiev, Golub, and
  Willander}}]{averkiev2:2002}
\bibinfo{author}{\bibfnamefont{N.~S.} \bibnamefont{Averkiev}},
  \bibinfo{author}{\bibfnamefont{L.~E.} \bibnamefont{Golub}}, \bibnamefont{and}
  \bibinfo{author}{\bibfnamefont{M.}~\bibnamefont{Willander}},
  \bibinfo{journal}{Semicond.} \textbf{\bibinfo{volume}{36}},
  \bibinfo{pages}{91} (\bibinfo{year}{2002}{\natexlab{b}}),
  \bibinfo{note}{[Fiz. Tekh. Poluprovodn. {\bf 36}, 97 (2002)]}.

\bibitem[{\citenamefont{Lau et~al.}(2001)\citenamefont{Lau, Olesberg, and
  Flatt\'e}}]{lau:2001}
\bibinfo{author}{\bibfnamefont{W.~H.} \bibnamefont{Lau}},
  \bibinfo{author}{\bibfnamefont{J.~T.} \bibnamefont{Olesberg}},
  \bibnamefont{and} \bibinfo{author}{\bibfnamefont{M.~E.}
  \bibnamefont{Flatt\'e}}, \bibinfo{journal}{Phys.\ Rev.~B}
  \textbf{\bibinfo{volume}{64}}, \bibinfo{pages}{161301(R)}
  (\bibinfo{year}{2001}).

\bibitem[{\citenamefont{Bir et~al.}(1976)\citenamefont{Bir, Aronov, and
  Pikus}}]{bir:1976}
\bibinfo{author}{\bibfnamefont{G.~L.} \bibnamefont{Bir}},
  \bibinfo{author}{\bibfnamefont{A.~G.} \bibnamefont{Aronov}},
  \bibnamefont{and} \bibinfo{author}{\bibfnamefont{G.~E.} \bibnamefont{Pikus}},
  \bibinfo{journal}{Sov. Phys. JETP} \textbf{\bibinfo{volume}{42}},
  \bibinfo{pages}{705} (\bibinfo{year}{1976}), \bibinfo{note}{[Zh. Eksp. Teor.
  Fiz. {\bf 69}, 1382 (1975)]}.

\bibitem[{\citenamefont{Dresselhaus}(1955)}]{dresselhaus:1955}
\bibinfo{author}{\bibfnamefont{G.}~\bibnamefont{Dresselhaus}},
  \bibinfo{journal}{Phys.\ Rev.} \textbf{\bibinfo{volume}{100}},
  \bibinfo{pages}{580} (\bibinfo{year}{1955}).

\bibitem[{\citenamefont{Bychkov and Rashba}(1984)}]{byc84}
\bibinfo{author}{\bibfnamefont{Y.~A.} \bibnamefont{Bychkov}} \bibnamefont{and}
  \bibinfo{author}{\bibfnamefont{E.~I.} \bibnamefont{Rashba}},
  \bibinfo{journal}{J.\ Phys.~C: Solid State Phys.}
  \textbf{\bibinfo{volume}{17}}, \bibinfo{pages}{6039} (\bibinfo{year}{1984}).

\bibitem[{\citenamefont{Rudolph et~al.}(2003)\citenamefont{Rudolph, H\"agele,
  Gibbs, Kithrova, and Oestreich}}]{rud03}
\bibinfo{author}{\bibfnamefont{J.}~\bibnamefont{Rudolph}},
  \bibinfo{author}{\bibfnamefont{D.}~\bibnamefont{H\"agele}},
  \bibinfo{author}{\bibfnamefont{H.~M.} \bibnamefont{Gibbs}},
  \bibinfo{author}{\bibfnamefont{G.}~\bibnamefont{Kithrova}}, \bibnamefont{and}
  \bibinfo{author}{\bibfnamefont{M.}~\bibnamefont{Oestreich}},
  \bibinfo{journal}{Appl.\ Phys.\ Lett.} \textbf{\bibinfo{volume}{82}},
  \bibinfo{pages}{4516} (\bibinfo{year}{2003}).

\bibitem[{\citenamefont{Winkler and R\"ossler}(1993)}]{winkler:1993}
\bibinfo{author}{\bibfnamefont{R.}~\bibnamefont{Winkler}} \bibnamefont{and}
  \bibinfo{author}{\bibfnamefont{U.}~\bibnamefont{R\"ossler}},
  \bibinfo{journal}{Phys.\ Rev.~B} \textbf{\bibinfo{volume}{48}},
  \bibinfo{pages}{8918} (\bibinfo{year}{1993}).

\bibitem[{\citenamefont{Terauchi et~al.}(1999)\citenamefont{Terauchi, Ohno,
  Adachi, Sato, Matsukara, Tackeuchi, and Ohno}}]{terauchi:1999}
\bibinfo{author}{\bibfnamefont{R.}~\bibnamefont{Terauchi}},
  \bibinfo{author}{\bibfnamefont{Y.}~\bibnamefont{Ohno}},
  \bibinfo{author}{\bibfnamefont{T.}~\bibnamefont{Adachi}},
  \bibinfo{author}{\bibfnamefont{A.}~\bibnamefont{Sato}},
  \bibinfo{author}{\bibfnamefont{F.}~\bibnamefont{Matsukara}},
  \bibinfo{author}{\bibfnamefont{A.}~\bibnamefont{Tackeuchi}},
  \bibnamefont{and} \bibinfo{author}{\bibfnamefont{H.}~\bibnamefont{Ohno}},
  \bibinfo{journal}{Jpn. J. Appl. Phys.} \textbf{\bibinfo{volume}{38}},
  \bibinfo{pages}{2549} (\bibinfo{year}{1999}).

\bibitem[{\citenamefont{Ohno et~al.}(2000)\citenamefont{Ohno, Terauchi, Adachi,
  Matsukura, and Ohno}}]{ohno:2000}
\bibinfo{author}{\bibfnamefont{Y.}~\bibnamefont{Ohno}},
  \bibinfo{author}{\bibfnamefont{R.}~\bibnamefont{Terauchi}},
  \bibinfo{author}{\bibfnamefont{T.}~\bibnamefont{Adachi}},
  \bibinfo{author}{\bibfnamefont{F.}~\bibnamefont{Matsukura}},
  \bibnamefont{and} \bibinfo{author}{\bibfnamefont{H.}~\bibnamefont{Ohno}},
  \bibinfo{journal}{Physica E} \textbf{\bibinfo{volume}{6}},
  \bibinfo{pages}{817} (\bibinfo{year}{2000}).

\bibitem[{\citenamefont{Glazov and Ivchenko}(2002)}]{glazov:2002}
\bibinfo{author}{\bibfnamefont{M.~M.} \bibnamefont{Glazov}} \bibnamefont{and}
  \bibinfo{author}{\bibfnamefont{E.~L.} \bibnamefont{Ivchenko}},
  \bibinfo{journal}{JETP Lett.} \textbf{\bibinfo{volume}{75}},
  \bibinfo{pages}{403} (\bibinfo{year}{2002}), \bibinfo{note}{[Pis'ma Zh. Eksp.
  Teor. Fiz. {\bf 75}, 476 (2002)]}.

\bibitem[{\citenamefont{Brand et~al.}(2002)\citenamefont{Brand, Malinowski,
  Karimov, Marsden, Harley, Shields, Sanvitto, Ritchie, and
  Simmons}}]{brand:2002}
\bibinfo{author}{\bibfnamefont{M.~A.} \bibnamefont{Brand}},
  \bibinfo{author}{\bibfnamefont{A.}~\bibnamefont{Malinowski}},
  \bibinfo{author}{\bibfnamefont{O.~Z.} \bibnamefont{Karimov}},
  \bibinfo{author}{\bibfnamefont{P.~A.} \bibnamefont{Marsden}},
  \bibinfo{author}{\bibfnamefont{R.~T.} \bibnamefont{Harley}},
  \bibinfo{author}{\bibfnamefont{A.~J.} \bibnamefont{Shields}},
  \bibinfo{author}{\bibfnamefont{D.}~\bibnamefont{Sanvitto}},
  \bibinfo{author}{\bibfnamefont{D.~A.} \bibnamefont{Ritchie}},
  \bibnamefont{and} \bibinfo{author}{\bibfnamefont{M.~Y.}
  \bibnamefont{Simmons}}, \bibinfo{journal}{Phys.\ Rev.\ Lett.}
  \textbf{\bibinfo{volume}{89}}, \bibinfo{pages}{236601}
  (\bibinfo{year}{2002}).

\bibitem[{\citenamefont{Braun and R\"ossler}(1985)}]{braun:1985}
\bibinfo{author}{\bibfnamefont{M.}~\bibnamefont{Braun}} \bibnamefont{and}
  \bibinfo{author}{\bibfnamefont{U.}~\bibnamefont{R\"ossler}},
  \bibinfo{journal}{J.\ Phys.~C: Solid State Phys.}
  \textbf{\bibinfo{volume}{18}}, \bibinfo{pages}{3365} (\bibinfo{year}{1985}).

\bibitem[{\citenamefont{Lundstrom}(2000)}]{lundstrom:2000}
\bibinfo{author}{\bibfnamefont{M.}~\bibnamefont{Lundstrom}},
  \emph{\bibinfo{title}{Fundamentals of Carrier Transport}}
  (\bibinfo{publisher}{Cambridge University Press},
  \bibinfo{address}{Cambridge}, \bibinfo{year}{2000}), \bibinfo{edition}{2nd}
  ed.

\bibitem[{del()}]{delta_mu}
\bibinfo{note}{This condition is fulfilled, if $\Delta \mu \ll k_B T$ or if
  $\Delta \mu \ll E_F$. In the latter case, $\hat{I}$ may only be applied to
  terms $E^\nu$ with exponents $\nu \ge 1$ and with $(\Delta\mu)^2 \ll 24 E_F^2
  / [\nu (\nu - 1)]$.}

\bibitem[{\citenamefont{Singh}(1993)}]{singh:1993}
\bibinfo{author}{\bibfnamefont{J.}~\bibnamefont{Singh}},
  \emph{\bibinfo{title}{Physics of Semiconductors and Their Heterostructures}}
  (\bibinfo{publisher}{McGraw-Hill}, \bibinfo{address}{New York},
  \bibinfo{year}{1993}).

\bibitem[{\citenamefont{Gradshteyn and Ryzhik}(1996)}]{gradshteyn:1996}
\bibinfo{author}{\bibfnamefont{I.~S.} \bibnamefont{Gradshteyn}}
  \bibnamefont{and} \bibinfo{author}{\bibfnamefont{I.~M.}
  \bibnamefont{Ryzhik}}, \emph{\bibinfo{title}{Table of Integrals, Series and
  Products}} (\bibinfo{publisher}{Academic}, \bibinfo{address}{San Francisco},
  \bibinfo{year}{1996}).

\bibitem[{res()}]{rescale}
\bibinfo{note}{Note that all rates are proportional to $\tau_\mathrm{tr}$ and
  can therefore be easily rescaled, e.\ g.\ to match measured values of
  $\tau_\mathrm{tr}$.}

\bibitem[{tau()}]{tau_z_def}
\bibinfo{note}{Due to different definitions of the spin relaxation time the
  values given here differ from those in Ref.\ \onlinecite{terauchi:1999} by a
  factor of two.}

\bibitem[{\citenamefont{Ohno}(2003)}]{ohno:2003}
\bibinfo{author}{\bibfnamefont{Y.}~\bibnamefont{Ohno}},
  \bibinfo{howpublished}{private communication} (\bibinfo{year}{2003}).

\bibitem[{\citenamefont{Glazov}(2003)}]{glazov:2003}
\bibinfo{author}{\bibfnamefont{M.~M.} \bibnamefont{Glazov}},
  \bibinfo{journal}{Phys. Solid State} \textbf{\bibinfo{volume}{45}},
  \bibinfo{pages}{1162} (\bibinfo{year}{2003}), \bibinfo{note}{[Fiz. Tverd.
  Tela, {\bf 45}, 1108 (2003)]}.

\bibitem[{\citenamefont{Glazov et~al.}(2003)\citenamefont{Glazov, Ivchenko,
  Brand, Karimov, and Harley}}]{glazov:cond-mat:2003}
\bibinfo{author}{\bibfnamefont{M.~M.} \bibnamefont{Glazov}},
  \bibinfo{author}{\bibfnamefont{E.~L.} \bibnamefont{Ivchenko}},
  \bibinfo{author}{\bibfnamefont{M.~A.} \bibnamefont{Brand}},
  \bibinfo{author}{\bibfnamefont{O.~Z.} \bibnamefont{Karimov}},
  \bibnamefont{and} \bibinfo{author}{\bibfnamefont{R.~T.}
  \bibnamefont{Harley}}, \bibinfo{journal}{cond-mat/0305260}
  (\bibinfo{year}{2003}), \bibinfo{note}{[11th Int. Symp. ``Nanostructures:
  Physics and Technology'', St. Petersburg, 2003, S. 273]}.

\bibitem[{\citenamefont{Oestreich}(2004)}]{oest04}
\bibinfo{author}{\bibfnamefont{M.}~\bibnamefont{Oestreich}},
  \bibinfo{howpublished}{private communication} (\bibinfo{year}{2004}).

\bibitem[{\citenamefont{Gantmakher and Levinson}(1987)}]{gantmakher:1987}
\bibinfo{author}{\bibfnamefont{V.~F.} \bibnamefont{Gantmakher}}
  \bibnamefont{and} \bibinfo{author}{\bibfnamefont{Y.~B.}
  \bibnamefont{Levinson}}, \emph{\bibinfo{title}{Carrier Scattering in Metals
  and Semiconductors}} (\bibinfo{publisher}{North-Holland},
  \bibinfo{address}{Amsterdam}, \bibinfo{year}{1987}).

\bibitem[{\citenamefont{Ridley}(1993)}]{ridley:1993}
\bibinfo{author}{\bibfnamefont{B.~K.} \bibnamefont{Ridley}},
  \emph{\bibinfo{title}{Quantum Processes in Semiconductors}}
  (\bibinfo{publisher}{Clarendon}, \bibinfo{address}{Oxford},
  \bibinfo{year}{1993}).

\bibitem[{\citenamefont{Ferreira and Bastard}(1989)}]{ferreira:1989}
\bibinfo{author}{\bibfnamefont{R.}~\bibnamefont{Ferreira}} \bibnamefont{and}
  \bibinfo{author}{\bibfnamefont{G.}~\bibnamefont{Bastard}},
  \bibinfo{journal}{Phys.\ Rev.~B} \textbf{\bibinfo{volume}{40}},
  \bibinfo{pages}{1074} (\bibinfo{year}{1989}).

\bibitem[{\citenamefont{Hess}(1979)}]{hess:1979}
\bibinfo{author}{\bibfnamefont{K.}~\bibnamefont{Hess}},
  \bibinfo{journal}{Appl.\ Phys.\ Lett.} \textbf{\bibinfo{volume}{35}},
  \bibinfo{pages}{484} (\bibinfo{year}{1979}).

\bibitem[{\citenamefont{Lee et~al.}(1983)\citenamefont{Lee, Shur, Drummond, and
  Morko\c{c}}}]{lee:1983}
\bibinfo{author}{\bibfnamefont{K.}~\bibnamefont{Lee}},
  \bibinfo{author}{\bibfnamefont{M.~S.} \bibnamefont{Shur}},
  \bibinfo{author}{\bibfnamefont{T.~J.} \bibnamefont{Drummond}},
  \bibnamefont{and}
  \bibinfo{author}{\bibfnamefont{H.}~\bibnamefont{Morko\c{c}}},
  \bibinfo{journal}{J.\ Appl.\ Phys.} \textbf{\bibinfo{volume}{54}},
  \bibinfo{pages}{6432} (\bibinfo{year}{1983}).

\bibitem[{\citenamefont{Pikus and Pikus}(1995)}]{pikus:1995}
\bibinfo{author}{\bibfnamefont{F.~G.} \bibnamefont{Pikus}} \bibnamefont{and}
  \bibinfo{author}{\bibfnamefont{G.~E.} \bibnamefont{Pikus}},
  \bibinfo{journal}{Phys.\ Rev.~B} \textbf{\bibinfo{volume}{51}},
  \bibinfo{pages}{16928} (\bibinfo{year}{1995}).

\end{thebibliography}
\end{document}